\documentclass[12pt]{article}
\usepackage{epsfig,latexsym,amsmath,amssymb,bm,cite}
\usepackage{graphicx}
\usepackage{color}
\newcommand{\be}{\begin{equation}}
\newcommand{\ee}{\end{equation}}
\newcommand{\bea}{\begin{eqnarray}}
\newcommand{\eea}{\end{eqnarray}}
\newcommand{\bem}{\begin{multline}}
\newcommand{\eem}{\end{multline}}
\newcommand{\beg}{\begin{gather}}
\newcommand{\eeg}{\end{gather}}

\newcommand{\ben}{\begin{eqnarray*}}
\newcommand{\een}{\end{eqnarray*}}

\begin{document}
\title{{\bf $q{\bar q}$ Potential at Finite T and Weak Coupling in ${\cal N}=4$ \\[1.5cm] }}
\author{
{\bf Anastasios Taliotis\thanks{e-mail: taliotis1@bohr.physics.arizona.edu, taliotis.1@osu.edu}}
\\[1cm] {\it\small Department of Physics, The University of Arizona}\\ 
{\it\small Tucson, AZ 85721, USA}\\[5mm]}

\date{November 2010}

\maketitle

\thispagestyle{empty}

\begin{abstract}
We compute the potential between a $q{\bar q}$ singlet for ${\cal N}=4$ $SUSY$ with gauge group $SU(N)$ at finite temperature $T$, large distances $rT \gg 1$ and weak coupling $g$. As a first step, we only consider at the electric modes and we compute the Debye mass $m_D$ where we find that each of the $8N$ bosonic degrees of freedom contributes to $m_D^2$ with $\frac{1}{6}g^2 T^2$ while each of the $8N$ fermionic degrees of freedom contributes with $\frac{1}{12}g^2 T^2$ yielding to $m_D^2=2Ng^2T^2$. Then, motivated by results obtained in the literature from both, a weak coupling approach for $QCD$ and a large coupling investigation of ${\cal N}=4$ $SUSY$ through $AdS/CFT$, we attempt to include magnetic mode corrections.
 \end{abstract}
\thispagestyle{empty}


\newpage

\tableofcontents

\setcounter{page}{1}

\section{Introduction}

Recently, the discovery of the $AdS/CFT$ correspondence \cite{Maldacena:1997re,Aharony:1999ti,Witten:1998qj,Gubser:1998bc} has created great interest in the ${\cal N}=4$ $SUSY$
 $SU(N)$ gauge theory  because the duality allows the investigation of several aspects of the gauge theory (see for instance \cite{Maldacena:1998im,Witten:1998zw,Kovchegov:2007pq,Brower:2010wf,Beuf:2009cx,Albacete:2008vs,Janik:2010we,Heller:2007qt,Kovchegov:2009du,Lin:2009pn,Janik:2005zt,Peschanski:2010cs,Gubser:2008pc,Gubser:2008yx,Albacete:2008dz,Iatrakis:2010jb,Gursoy:2010fj,Gursoy:2009kk,Kovchegov:2010zg,Cornalba:2010vk,Cornalba:2009ax,Taliotis:2010pi,Lin:2010cb,Athanasiou:2010pv,D'Eramo:2010ak,Dumitru:2009ni,Noronha:2009ia}) at strong coupling through gravitational and generally classical computations.

In particular one of the first computations using the gauge/gravity duality was the evaluation of the heavy-quark potential at large 't Hooft coupling (and large $N$), both at zero temperature \cite{Maldacena:1998im} and also at finite temperature \cite{Rey:1998bq,Brandhuber:1998bs,Albacete:2008dz,Albacete:2009bp}. In particular, in \cite{Albacete:2008dz,Albacete:2009bp} it was found that at large enough distances the real part of the potential has a power low fall-off extending the results (obtained ealrier) in the literature \cite{Rey:1998bq,Brandhuber:1998bs}. The method of \cite{Albacete:2008dz,Albacete:2009bp} involved an analytic continuation of the solution of \cite{Rey:1998bq,Brandhuber:1998bs} applying the ideas of \cite{Albacete:2008ze,Taliotis:2009ne,Albacete:2008vv}.

Amazingly enough, exactly the same power fall-off was found in \cite{Gale:1987en} for pQCD at finite temperature. This fact partially motivated investigating the ${\cal N}=4$ $SUSY$ $SU(N)$ gauge theory at non-zero temperature in the framework of a pertubative, field theoretical approach.

Therefore, in this work we compute the heavy-quark singlet potential of ${\cal N}=4$ $SUSY$ at non-zero temperature at weak 't Hooft coupling. We compute the Debye mass and apply the technique of \cite{Gale:1987en} that had been previously applied to QCD, in our case, and investigate whether we may predict a similar power low fall-off of the potential. Such an investigation allows for the comparison with the result of \cite{Gale:1987en} obtained in the framework of pQCD and with the result of \cite{Albacete:2008dz,Albacete:2009bp} obtain in the framework of $AdS/CFT$. We note that literature is reach in regards of perturbative calculations at finite temperature for both, the heavy-quark potential for $QCD$ \cite{Kuti:1980gh,Dumitru:2010id,Gale:1987en,Kajantie:1982xx,DeGrand:1986uf} as well as several (other) contexts of the ${\cal N}=4$ $SUSY$ theory \cite{Chesler:2006gr,Fotopoulos:1998es,KorthalsAltes:2009dp,Blaizot:2006tk,Yamada:2006rx}.
 
\vspace{0.1in}
We organize this work as follows: 
\vspace{0.1in}

In section \ref{SU} we set up the problem arguing that the singlet potential is a gauge invariant quantity and hence it makes sense to compute (perturbatively).  We also give an elementary introduction of the objects we need in this work and which appear in thermal field theory.

Section \ref{diag} deals with the relevant diagrams involved in the calculation. For a subset of the diagrams under consideration, we borrow results from the literature calculated for $QCD$ and we show how these results may be adapted for our case.

In section \ref{poten} we use the results of previous sections in order to evaluate the quark-pair potential. We initially compute the Debye mass giving rise to the expected Yukawa fall off and then we extend the computation in the spirit of \cite{Gale:1987en}.

Finally, in section \ref{conc}  we summarize and discuss our results. In particular, we find that the potential has a power fall-off (at sufficiently large distances) which agrees precisely with \cite{Gale:1987en} and (the absolute value of) \cite{Albacete:2008dz,Albacete:2009bp}.

The notation and conventions we use as well as many useful formulas may be found in Appendix \ref{A}. In the rest of the Appendices we explicitly write the Lagranian for ${\cal N}=4$ $SUSY$ and derive the Feynman Rules. We also show which diagrams contribute in the calculation we are interested and evaluate in great detail (one of) the relevant diagram(s) in order to exhibit the general idea behind these sorts of calculations. 

\section{Setting up the Problem}\label{SU}

\subsection{$q{\bar q}$ potential and gauge invariance}\label{GI}

In $QED$, the magnetic and electric fields are gauge invariant. In non-abelian gauge theories on the other hand, gauge invariance (of the chromo-electromagnetic field) is not valid as a consequence of the presence of the non-linear terms in the field strength tensor $F_{\mu \nu}^a$. However, one gauge invariant quantity one may construct is the free energy (potential) of a static, color-singlet (total color charge zero), quark-antiquark pair as a function of the separation $r$ of the pair \cite{Kapusta:2006pm}. As the quantity we wish to compute is gauge invariant, we choose to compute it in the Temporal Axial Gauge ({\bf TAG}).


\subsection{Elements of thermal field theory}

In this subsection we present the minimum knowledge about field theory at finite temperature one needs in order to perform the calculation we are interested. 

The self energy of the gauge boson (2-point irreducible diagram\footnote{Whose precise definition is given below in equation (\ref{Dexact}).}) at finite temperature may be decomposed as
\begin{align}\label{gse}
\Pi^{\mu \nu}=G(q^0,|{\bf q}|)P_T^{\mu \nu}+F(q^0,|{\bf q}|)P_L^{\mu \nu}
\end{align} 
where $q^{\mu}$ is the four-momentum of the gauge boson, $G$ and $F$ are scalar functions of $q^0$ and $|{\bf q}|$ while $P_T$ and $P_L$ are the transverse and longitudinal projection tensors (to ${\bf q}$) respectively and their explicit form is given by 
\begin{align}\label{ptl}
P_T^{\mu \nu}=  \left( \delta_{ij}-\frac{q_iq_j}{{\bf q}^2}\right) g^{i\mu}g^{j\nu}              \hspace{0.3in} P_L^{\mu \nu}=(-g^{\mu \nu}+\frac{q^{\nu}q^{\mu}}{q^2})-P_T^{\mu \nu}, \hspace{0.1in} \mu={0,i}.
\end{align} 
These tensors satisfy
\begin{align}\label{proj}
q_iP_T^{i \nu}=q_{\mu}P_T^{\mu \nu}=q_{\mu}P_L^{\mu \nu}=P_T^{\mu \kappa}P_{L \hspace{0.01in} \kappa}\hspace{0.01in}^{ \nu}=0.
\end{align} 
Motivated by equations (\ref{proj}) and (\ref{ptl}), it is natural to associate $G$ with the chromomagnetic modes while $F$ with the chromoelectric one's. These factors may be expressed in terms of the diagonal components of $\Pi^{\mu \nu}$ as
\begin{align}\label{GF}
F=\frac{k^2}{{\bf k}^2}\Pi^{00} \hspace{0.3in} F+2G=\Pi^{ii}-\Pi^{00}.
\end{align} 
Taking into account the expression for the bare propagator of the gluon $D^{\mu \nu}_0$ from figure \ref{propFR} written in the TAG gauge, one may formally at least, compute the exact (dressed) propagator to all orders in perturbation theory whose non zero components are  
\begin{align}\label{Dexact}
D^{i j}=D_0^{i k} \sum_{N=0}^{\infty}[(-\Pi D_0)^N]_{k}\hspace{0.015in}^{j}=\frac{1}{G-k^2} \left( \delta_{ij}-\frac{q_iq_j}{{\bf q}^2}\right) +\frac{1}{F-k^2}\frac{k^2}{k_0^2}\frac{k^ik^j}{{\bf k}^2} \hspace{0.15in}  \mathrm{(TAG)}.
\end{align} 
This expression in fact provides the definition for $\Pi^{\mu \nu}$.

Now let us suppose that we place two oppositely (cromo)charged fermions at a distance $r$ and seek for the mutual force of the two charges treating them as small (static) perturbations in the thermal medium.  Then, in the spirit of the linear response theory one may show \cite{Kapusta:2006pm} that the potential of the $q{\bar q}$
system as a function of the separation is given by 
\begin{align}\label{qqV}
V(r)=Q_1Q_2\int \frac{d^3q}{(2 \pi)^3}\frac{e^{i {\bf k}{\bf r}}}{{\bf q}^2+F(0,{\bf q})}=Q_1Q_2\int \frac{d^3q}{(2 \pi)^3}\frac{e^{i {\bf k}{\bf r}}}{{\bf q}^2-\Pi^{00}(0,{\bf q})}.
\end{align} 
where the last equality is a consequence of the (left) equation (\ref{GF}) and (\ref{gmn}). Thus, we conclude that in order to compute the potential we only need the (00) component of $\Pi^{\mu \nu}$. In this work, we compute it perturbatively to $O(g^2T^2)$ (see section \ref{diag}). 

\subsection{Approximations}

As we will see, $\Pi^{00}$ depends from the temperature $T$ and the chemical potential $\mu_i$\footnote{The index $i$ is associated with every global symmetry $(i)$ that gives rise to $\mu_i$.} through the ratio $\mu_i/T$, that is
\begin{align}\label{PiTmu}
\Pi^{00}=\Pi^{00}(q^0,{\bf q};T, \mu_i/T).
\end{align} 
 We will work in the region where
\begin{align}\label{approx}
\frac{1}{\mu_i} \gg \frac{1}{T} \hspace{0,1in}\forall \hspace{0,015in} i    \hspace {0.6in}  r\gg \frac{1}{T}  \hspace{0.015in}.
\end{align} 
In particular, the right approximation above implies that the integral in equation (\ref{qqV}) receives its main contribution from ${\bf q} \rightarrow 0 $ and as a result (\ref{qqV}) reduces to 
\begin{align}\label{appV}
V(r)=Q_1Q_2\int \frac{d^3q}{(2 \pi)^3}\frac{e^{i {\bf k}{\bf r}}}{{\bf q}^2-  \lim \limits_{{\bf q} \to 0} \Pi^{00}(0,{\bf q})}.
\end{align} 
where in the color singlet state, the product of charges is
\begin{align}\label{q1q2}
Q_1Q_2=(gT^a_G) \otimes (-g T^a_{{\bar G}}) =-C_2(G) g^2=-N g^2.
\end{align} 
The expression $T^a_G \otimes T^a_{{\bar G}}$ is the tensor product of the adjoint times the (anti)adjoint
representation resulting to the Casimir operator and which in turn yields to the expected factor of $N$.

\section{Evaluating the Relevant Diagrams} \label{diag}
\begin{figure}[h!]
\centering
\includegraphics*[scale=0.7,angle=0,clip=true]{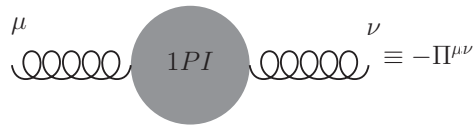} \hspace{1cm}
\caption{Diagramatic definition of (the colored) $\Pi^{\mu \nu}$ suppressing color indices. There is no imaginary $i$ involved in the definition.}
\label{1PI}
\end{figure}

The gauge boson self energy tensor is defined diagrammatically through figure \ref{1PI}.
%
The diagrams one has to calculate in order to compute $V(r)$ to $O(g^2)$ are given in Appendix \ref{C}. Most of them have been calculated and may be found (in addition to other places) in \cite{Kapusta:2006pm}. The important point here is that there is not any scalar propagator involved in this computation because according to subsection \ref{GI} the potential we compute involves a color singlet and hence the fermion pair should be of the same flavor. However, according to the Lagranian (\ref{LN4}) and the resulting Feynman rules for scalars, figures \ref{Yuk}, the vertices change the flavor of the fermions (observe the presence $\epsilon_{ijk}$ in the corresponding Feynman rules) concluding that scalar propagators do not contribute to the calculation under consideration.\footnote{In practice one may imagine that the two interacting fermions are massive, charged under color and couple (directly) only to the gauge bosons but not to the scalars. It is evident that these fermions are external to the ${\cal N}=4$ $SUSY$ particles.}

\subsection{Contribution of $\Pi^{00}_g$ due to the gauge loops}

The contribution due to the gluons arise from the diagrams of figure \ref{pig}. The result in the TAG is given by equation (8.65) of \cite{Kapusta:2006pm} and may also be found in \cite{Kajantie:1982xx} and is given by 
\begin{align}\label{p00g}
\Pi^{00;ab}_{g \hspace{0.01in}mat}(q_0,|{\bf q}|)=&-\delta^{ab} \frac{g^2N}{4\pi^2} \int_0^{\infty} dk k N_B(k) \times 
\mbox{Re} \Bigg[4-\frac{({\bf q}^2-2kq_0-q_0^2)(2k+q_0)^2}{2k^2(k+q_0)^2}+\frac{(2k+q_0)^2}{2k|{\bf q}|}\times\notag\\&
 \left(1+\frac{(k^2+(k+q_0)^2-{\bf q}^2)^2}{4k^2(k+q_0)^2})\ln \left( \frac{R_{g+}}{R_{g-}} \right) \right) \Bigg], \hspace{0.3in} q_0=i 2\pi n T, \hspace{0.015in}k\equiv|{\bf k}|
\end{align} 
where $N_B(k)$ is defined in (\ref{NFB}), $R_{g\pm}={\bf q}^2-2kq_0-q_0^2\pm 2k|{\bf q}| $ 
while the operator Re is defined in (\ref{Re}). In the limits we are interested (see (\ref{approx})) we eventually obtain that the matter (thermal) part contribution is
\begin{align}\label{p00gL}
\Pi^{00;ab}_{g \hspace{0.01in}mat}(0,|{\bf q}|\rightarrow 0)= -\delta^{ab} Ng^2\left(\frac{1}{3}T^2-\frac{1}{4}T |{\bf q}| +O({\bf q}^2\log({\bf q}^2/T^2))\right), \hspace{0.2in} q\ll T
\end{align} 
where we have used the left equation of (\ref{IBF}) and (\ref{I4}) in order to extract the zeroth and first order in $|{\bf q}|$ respectively.

\subsection{Contribution of $\Pi^{00}_f$ due to the fermion loops}

This is the contribution due to the diagrams of figure \ref{pif}. For a Dirac fermion this contribution has been calculated in the literature and has been reproduced by us.  For instance in \cite{Kapusta:2006pm}, equation (5.51), gives the answer (matter contribution) for an abelian gauge theory ($QED$). In order to adapt the answer to our case, we must multiply (5.51) by $f^{acd}f^{acd}=\delta^{ab} N$ to account for the color factors times $4$\footnote{Assuming they have the same chemical potentials $\mu_i$ for simplicity; in any case the $\mu_i$'s will not contribute (to zeroth order) in the view of (\ref{approx}).}
to account for the four fermions (one gaugino and three adjoint fermions; see (\ref{LN4})). Finally, we should multiply by $(1/2)$ in order to account for the fact that we deal with Weyl spinors whose corresponding trace involves (four) sigma (instead of gamama) matrices; in the view of (\ref{traces}) the overall factor is $2$ instead of $4$ which occurs with the gamma matrices. The antisymmetric part in the trace cancels because $\Pi^{00}$ is symmetric in the spacetime indices. Therefore, we effectively multiply equation (5.51) of \cite{Kapusta:2006pm} by $2N\delta^{ab}$ assuming massless fermions yielding to
\begin{align}\label{p00f}
\Pi^{00;ab}_{f \hspace{0.01in} mat}(q_0,|{\bf q}|)=&-2\delta^{ab} \frac{g^2N}{\pi^2} \mbox{Re} \int_0^{\infty} dk k (N^+_F(k)+N^-_F(k)) \times \notag\\&
 \Bigg[1 + \frac{4kq_0-4k^2-q_0^2+q^2}{4 k |{\bf q}|}  \ln \left( \frac{R_{f+}}{R_{f-}} \right) \Bigg], 
 \hspace{0.3in} q_0=i 2\pi n T
\end{align} 
where $R_{f \pm}=q_o^2-{\bf q}^2-2q_0k\pm 2k|{\bf q}|$. In the limits that we are concerned  (see (\ref{approx}) and (\ref{appV})) we obtain that the matter (thermal) part is

\begin{align}\label{p00fL}
\Pi^{00;ab}_{f \hspace{0.01in}mat}(0,|{\bf q}|\rightarrow0)&=-4\delta^{ab} \frac{g^2N}{\pi^2}  \int_0^{\infty} dk k N_F(k) \Bigg[1 - \frac{k}{ |{\bf q}|}  \ln \left( \frac{R_{f+}(q^0=0)}{R_{f-}(q^0=0)} \right) \Bigg]  +O({\bf q}^2)  \notag\\&
=-\frac{2}{3}\delta^{ab}Ng^2T^2+O({\bf q}^2)
\end{align} 
where  in extracting the zeroth contribution to $|{\bf q}|$ we have used (the right integral of) (\ref{IBF}). We point out that there is not a linear term in $q^{\mu}$ present unlike (\ref{p00gL}) and that in extracting the leading terms in $|{\bf q}|$ one should be careful (see (\ref{I3}) for one example).

\subsection{Contribution of $\Pi^{00}_{s_i}$ due to the scalar loops}

The corresponding diagrams are the two diagrams of figure \ref{pis}. 

{\bf First Contribution $(s_1)$:} This is due to the diagram on the left of the figure and has been computed in the literature for the case of $\phi^4$ theory (see equation (3.48) of  \cite{Kapusta:2006pm}). In order to adapt it for our case we should, according to the Feynman rule of figure \ref{gsFR} (in the middle), replace $\lambda \rightarrow g^2 \left(f^{cad}f^{dbe}+f^{cbd}f^{dae} \right)\delta^{ce} \delta_{ii} g^{00}=-6g^2N\delta^{ab}$ as expected.\footnote{A factor of $2$ exists in scalar QED as well while $3N$ shows all the different complex colored scalars in the loop. The additional minus sign is traced simply to the fact that instead of $-\lambda$ it is used $g^2$. The chemical potentials are again ignored.}We find that the contribution of this diagram is
\begin{align}\label{p00s1L}
\Pi^{00;ab}_{s_1 \hspace{0.01in}mat} (q_0,|{\bf q}|)= 6g^2 N  \delta^{ab}T\sum_n \int \frac{d^3{\bf k}}{(2 \pi)^3}\frac{1}{(k^0_n)^2-k^2}  =  
-6 g^2N \delta^{ab}\int \frac{d^3{\bf k}}{(2 \pi)^3}\frac{1}{k} N_B(k) =-\delta^{ab} \frac{1}{2}g^2NT^2
\end{align} 
where $k_n^0= i 2\pi nT$, $ n \in \mathbb{Z}$, $k \equiv |{\bf k}|$ and $N_B$ is given by (\ref{NFB}). The first equality is a direct application of the Feynman rules for thermal field theory, the second equality follows by integrating using (\ref{sumB}) and the third equality uses the left integral of (\ref{IBF}). We note that this diagram is $q^{\mu}$ independent. 

{\bf Second Contribution $(s_2)$:} This is due to the diagram on the right of figure \ref{pis} and which we evaluate in the Appendix \ref{D}\footnote{We evaluate this diagram in great detail (also ignoring the chemical potential contribution) as the rest of the five diagrams (see (\ref{p00gL}), (\ref{p00fL}) and (\ref{p00s1L})), are evaluated in an analogous way.} obtaining equation (\ref{sd}). For $q^0=0$ which is what interests us we have  
\begin{align}\label{p00s2}
\Pi^{00;ab}_{s_2 \hspace{0.01in}mat}(q_0,|{\bf q}|)=- 3\frac{g^2}{4 \pi^2} \delta^{ab}   \int _0^{\infty}  dx \frac{1}{2}x^2 |{\bf q}|^2  \ln \left( \frac{1+x}{|1-x|}\right) N_B(\frac{|{\bf q}|}{2} x)
\end{align} 
%
where we have made for the integral (\ref{sd}) the substitution $2k=|{\bf q}| x$. Extracting the leading and sub-leading contribution for small $ |{\bf q}|$ is not straightforward but it may be achieved by working in the same way as in proving (\ref{I3}) yielding
\begin{align}\label{p00s2L}
\Pi^{00;ab}_{s_2 \hspace{0.01in}mat}(0,|{\bf q}| \rightarrow 0)=-\frac{1}{2} \delta^{ab} N g^2 T^2+O( |{\bf q}|^2).
\end{align} 
It is crucial to highlight that likewise the fermion case, the scalars have no linear contribution in $ |{\bf q}|$ either. This completes the set of contributions to the order in $g$ and $ |{\bf q}|$ that we are interested.
\section{Calculating the Potential}\label{poten}

Collecting the results from equations (\ref{p00gL}), (\ref{p00fL}), \ref{p00s1L}) and (\ref{p00s2L}) the self energy tensor to $O(g^2)$ is
\begin{align}\label{P00}
\Pi^{00}_{mat}(q_0=0,|{\bf q}\rightarrow 0| )= -m_D^2+2 m_D t|{\bf q}| +O({\bf q}^2\log({\bf q}^2/T^2))
\end{align} 
where $m_D$ is the Debye mass and is given by
\begin{align}\label{mD}
m_D=2Ng^2 T^2.
\end{align} 
This is one of our main results  for this work. On the other hand $t$ is positive and is given by
\begin{align}\label{t}
t=\frac{1}{16 } \frac{m_D}{T}.
\end{align}
\subsection{$V(r)$ from pure electric mode corrections}
Performing the integral of (\ref{appV}) using just the $m_D^2$ for $\Pi^{00}$ we obtain
\begin{align}\label{YukV}
V(r)=-\frac{g^2 N}{4 \pi} \frac{1}{r} e^{-m_D r}
\end{align} 
which is the expected Yukawa potential.
\subsection{$V(r)$ from magnetic mode corrections}
One may push the calculation a little further and include corrections to the potential due to the linear contribution of $|{\bf q}|$ in $\Pi(0,{\bf q})$. As this term is gauge invariant \cite{Toimela:1982hv,Gale:1987en} it is tempting to try to include it in the evaluation the potential extending the result of (\ref{YukV}). Although it has not been rigorously proved that this expansion in $|{\bf q}|$ is well defined, it is believed \cite{Gale:1987en} that it is very likely to be the case. Assuming this and applying the analysis of \cite{Gale:1987en} for our case, we find that 
\begin{align}\label{V1}
V(r)=\frac{2 g^2N }{\pi^2} \frac{t}{m_D^3 r^4}=\frac{1}{(4 \pi)^2} \frac{1}{T^3 r^4}
\end{align} 
which shows that at sufficiently large distances the potential falls of as $1/r^4$ and also it is repulsive and in addition $N$ and $g$ independent! This is the second main part of our investigation.
\section{Discussion}\label{conc}
In this work we perform the calculation of the $q{\bar q}$ potential in a thermal medium for the ${\cal N}=4$ $SUSY$ theory at weak coupling. By considering the purely electric modes at high temperatures we find the expected Yukawa potential, equation (\ref{YukV}), with the Debye mass given by equation (\ref{mD}). In particular, we observe that each of the (eight $\times N$) bosonic degrees of freedom  contribute to $m_D^2$ with $\frac{1}{6}g^2T^2$ (see (\ref{p00g}), (\ref{p00s1L}) and (\ref{p00s2L})) while each of the (eight $\times N$) fermionic degrees of freedom contributes with $\frac{1}{12}g^2T^2$ (see (\ref{p00fL})) leading to (\ref{mD}). 
 
Next, and following \cite{Gale:1987en}, we include (a subset of the) magnetic corrections obtaining the potential of equation (\ref{V1}) at large distances which is independent from the coupling and the number of colors and falls of as $1/r^4$ but it is repulsive (as in \cite{Gale:1987en}). On the other hand, from $AdS/CFT$ calculations it was found  \cite{Albacete:2008dz}  the same power fall-off at large distances but with an attractive force between the $q{\bar q}$. Motivated by that result, we were hoping, by including the scalars contribution for the ${\cal N}=4$ $SUSY$ theory at weak coupling, to obtain the result of \cite{Gale:1987en} with an additional overall negative sign agreeing with \cite{Albacete:2008dz} and also with our intuition. We find that both the fermions and the scalars do not contribute to the magnetic modes (linearly in ${\bf |q|}$) and hence the analysis of  \cite{Gale:1987en} proceeds (up to an overall factor) unaltered yielding the repulsive potential of equation (\ref{V1}).

This work provides a concrete example of an observable of ${\cal N}=4$ $SUSY$ whose behavior is qualitatively the same as the corresponding one of $QCD$. Consequently one may conclude that studying non perturbative phenomena of $QCD$ by applying the $AdS/CFT$ correspondence may not be far from reality and has the potential to yield to qualitatively correct results.
\section*{Acknowledgments} 
We would like to thank A. Majumder for  fruitful discussions and also for giving a great set of lectures on Field Theory at Finite Temperature in spring of 2010 at The Ohio State University. We  acknowledge S. Dam for useful discussions during TASI School 2010 which has been inspiring. We also thank the organizers of ``Quantifying the Properties of Hot QCD Matter" and especially U. Heinz, held at INT in Seattle, the organizers of HESI10 and especially K. Fukushima held at the Yukawa Institute in Kyoto and the organizers of  the ``AdS Holography and the QGP"/the EMMI workshop for ``Hot Matter", and especially A. Rebhan,  held at ESI in Vienna, for their excellent hospitality during the set-up of this project. Finally, we would like to thank E. Braaten, H. Haber, I. Sarcevic, K. Dienes, John Terning, A. Stergiou, J. Kapusta and Y. Kovchegov for informative discussions, S. Avery for reading the manuscript and making extremely useful comments and in particular Y. Kovchegov for sharing his striking insight. 

\vspace{0.1in}
This work is supported by The Department of Energy under contracts DE-FG02-04ER41319 and DE-FG02-04ER41298 and partially by the Institute of Governmental Scholarships of Cyprus.

 


\appendix
\renewcommand{\theequation}{A\arabic{equation}}
\setcounter{equation}{0}
\section{Conventions, Notation and Useful Formulas}
\label{A}
The metric tensor we use has signature
\begin{align}\label{gmn}
g_{\mu\nu}=(1,-1,-1,-1).
\end{align} 
For 4-vectors we use the standard notation
\begin{align}\label{kokvec}
k^{\mu}=(k^0,{\bf k})=(k^0,k^i)  \hspace{0.3in} \mu=0;1,2,3=0;i.
\end{align} 
We define the $\beta$ factor to be inversely proportional to the temperature T
\begin{align}\label{beta}
\beta=\frac{1}{T}.
\end{align} 
The generators of $SU(N)$, $T^a$, in the fundamental representation are normalized as
\begin{align}\label{normT}
Tr[T^aT^b]=\frac{1}{2}\delta^{ab}, \hspace{0.1in}a,b=1,2,...N.
\end{align} 
and they obey the algebra
\begin{align}\label{alg}
[T^a,T^b] = if^{abc}T^c\hspace{0.3in}  \mbox{where $f^{abc}$ are the structure constants}.
\end{align} 
The adjoint representation is defined by
\begin{align}\label{adj}
(T^b)_{ac} = if^{abc}.
\end{align} 
The following summation formula for the structure constants is valid
\begin{align}\label{sff}
f^{acd}f^{bcd}=N \delta^{ab}.
\end{align} 
The covariant derivative in the fundamental and the adjoint representation for {\bf negative charge g},\footnote{This convention agrees with \cite{Terning:2006bq} (see eq. (2.96)), \cite{Dreiner:2008tw}  (see eq. (4.3.10)) and \cite{Wess:1992cp} (see exercise 7, Ch. $VII$).} that is for charge $-|g|=-g$ are defined by
\begin{align}\label{covder}
D_\mu=\mathbb{I}_{N}\partial_{\mu}+igA^a_{\mu}T^a \hspace{0.3in}\delta^{ac}D_\mu=\delta^{ac}\partial_{\mu}-gA^b_{\mu}f^{abc} .
\end{align} 
We define the Pauli matrices in a four-vector form with doted and un-doted indices as in \cite{Dreiner:2008tw}
\begin{align}\label{sigma}
\sigma^{\mu}_{ \alpha {\dot \alpha}}=(\mathbb{I}_2,{\bf \sigma})  \hspace{0.3in}  {\bar \sigma}^{\mu \hspace{0.015in  {\dot \alpha} \alpha}}=(\mathbb{I}_2,-{\bf \sigma})
\end{align} 
The trace formulas for four (alternate) sigma matrices are
\begin{subequations}\label{traces}
\begin{align}
&Tr[\sigma^{\mu}{\bar \sigma}^{\nu}\sigma^{\rho}\bar{\sigma}^{\kappa}]=  2\left( g^{\mu \nu}
g^{\rho \kappa}- g^{\mu \rho}g^{\nu \kappa}+ g^{\mu \kappa}g^{\nu\rho}+i\epsilon^{\mu \nu \rho \kappa}\right), \label{tr1}  \\ 
&
Tr[{\bar \sigma}^{\mu}\sigma^{\nu}\bar{\sigma}^{\rho} \sigma^{\kappa}]=  2\left( g^{\mu \nu}
g^{\rho \kappa}- g^{\mu \rho}g^{\nu \kappa}+ g^{\mu \kappa}g^{\nu\rho}-i\epsilon^{\mu \nu \rho \kappa}\right). \label{tra2}
\hspace{0.3in} \end{align} 
\end{subequations}
We define the fermion and the boson distributions in the presence of a chemical potential (which is a consequence of some global symmetry) by
\begin{subequations}
\begin{align}\label{NFB}
&N_F^{\pm}(p)=\frac{1}{\mathrm{exp}[\beta({E_p\pm \mu_F})]+1} \hspace{0.3in} N_B^{\pm}(p)=\frac{1}{\mathrm{exp}[\beta({E_p\pm \mu_B})]-1}, \hspace{0.15in}\beta=1/T \\&
 N_F(p)=N_F^{\pm}(\mu_F=0,p) \hspace{0.75in}N_B(p)=N_B^{\pm}(\mu_B=0,p). \label{NFBb}
\end{align} 
\end{subequations}
We will also need the summation formulas

\begin{subequations}\begin{align}\label{sumfo}
&T\sum_{n}f(p^0=i\omega_n+\mu)=\pm \frac{1}{2 \pi i}\int_{-i\infty+\mu+\epsilon}^{i\infty+\mu+\epsilon}  dp^0 \frac{f(p^0)}{e^{\beta(p^0-\mu)}\mp1}           \pm \frac{1}{2 \pi i}\int_{-i\infty+\mu-\epsilon}^{i\infty+\mu-\epsilon}  dp^0 \frac{f(p^0)}{e^{-\beta(p^0-\mu)}\mp1}     \notag\\&
\hspace{1.7in}+\frac{1}{2 \pi i}\int_{-i\infty}^{i\infty}  dp^0 f(p^0)+\frac{1}{2 \pi i}\oint_C  dp^0 f(p^0), \hspace{0.1in}\epsilon>0,\\&
\hspace{0.6in} \omega_n=\frac{2\pi}{\beta} n \hspace{0.1in} \mbox{for bosonic modes}  \hspace{0.5in} \omega_n=\frac{2\pi}{\beta} (2n+1) \hspace{0.1in} \mbox{for fermionic modes} 
\end{align} 
\end{subequations}
and where $C$ is a closed rectangular contour with long sides along the real axis at $p^0=0$ and $p^0=\mu$ which extend from $i\infty$ to $-i\infty$ and from $-i\infty+\mu$ to $+i\infty+\mu$ respectively. The contour closes at infinities and it evidently has a counter-clock-wise direction. We note that the chemical potential $\mu$ is not generally the same for bosons and fermions and hence an extra index is (generally) required. The first two terms of (\ref{sumfo}) show the thermal contribution of the particle-antiparticle and they vanish at $T=0$, the third term is the vacuum piece at $T=0$ just as in ordinary field theory while the last term  is the contribution of the matter at finite density but at $T=0$. 

It is useful to define
\begin{align}\label{Re}
\mbox{Re} \left[f(k_0,{\bf k},x) \right]  \equiv \frac{1}{2} \left(f(k_0,{\bf k},x)+f(-k_0,{\bf k},x) \right), \hspace{0.015in} \mbox{for any set of parameters x.}
\end{align} 
Finally, we will need the integrals
\begin{subequations}\begin{align}\label{I1}
&\hspace{0.7in} \int_{0}^{\infty}dp \frac{p}{\mbox{exp}[\beta(p-\mu)]\mp 1}=\pm T^2 Li_2(\pm e^{\mu/T}) \\&
\int_{0}^{\infty}dp \frac{p}{\mbox{exp}(\beta p)- 1}= \frac{\pi^2T^2}{6}  \hspace{0.3in}  \int_{0}^{\infty}dp \frac{p}{\mbox{exp}(\beta p)+ 1}= \frac{\pi^2T^2}{12} \label{IBF}
\end{align} 
\end{subequations}
where $ Li_2$ is the dilogarithm function. We also need the asymptotic expansions
\begin{subequations}
\begin{align}\label{I3}
& \int_{0}^{\infty}dx \frac{1}{e^{a x}-1}\ln \left( \frac{1+x}{|1-x|} \right)=\frac{\pi^2}{2a}+\ln \left( \frac{a}{2 \pi e} \right) +\gamma_{E}+O(a),
 \\&
 \int_{0}^{\infty}dk \frac{1}{e^{ k/T}-1}\ln \left( \frac{q+2k}{|q-2k|} \right)=\frac{1}{2}\pi^2 T+\frac{1}{2}q \ln \left( \frac{q}{4 \pi e T} \right) +\frac{1}{2}\gamma_E q+O(q^2)
 \label{I4}
\end{align} 
\end{subequations}
where $\gamma_E$ is the Euler constant while (\ref{I4}) follows directly from (\ref{I3}). We sketch the proof of (\ref{I3}) below.

{\bf Proof of (\ref{I3}):} Break the interval of integration in the intervals $I_1=(0,1)$ and $I_2=(1,\infty)$. For the interval $I_1$, expand the logarithm in powers of $x$ and also expand $(e^{ax}-1)^{-1}=1/ax-1/2+O(ax)$ and multiply it with the series resulting from the logarithm and perform the integrations recognizing singular and zeroth terms in $a$. Resum the (two resulting) series to obtain $I_1=\pi^2/2a-\ln(2)$. For $I_2$ expand the logarithm in inverse powers of x and make the transformation $ax \rightarrow x$. This makes the integration range $(1,\infty) \rightarrow (a, \infty)$. Break this new integration region from $I_2^b=(a,1)$ to $I_2^a=(1,\infty)$. The $I_2^a$ range gives only a zeroth order contribution which is given by $I_2^a=2\int_{1}^{\infty}\frac{dx}{x} \frac{1}{e^{ x}-1}$. For the  $I_2^b$ range expand  $(e^{ax}-1)^{-1}=c_{-1}/ax+c_0+\sum_{n\geq 1} c_n x^n$ for appropriate coefficients $c_n$ with $c_{-1}=1$ and $c_0=-1/2$. The inverse powers of $a$ are in the $c_{-1}$ for which when the corresponding series (due to the logarithm) is resumed, gives $\pi^2/4a-2$. The $log(a)$ term comes from $c_0$ where there is no series to be resumed. The zeroth orders in $a$ come either from $c_0$ and an appropriate series which yields to $\ln(2)-1$ or from $2\sum_{n\geq 1} \int_{0}^{1}c_n x^{n-1}=\sum_{n\geq 1} c_n/n$. This series may be written as $2\int_{0}^{1} \frac{dx}{x} (\frac{1}{e^{x}-1}-1/x+1/2) $. Add the pieces to obtain $I=I_1+I_2=I_1+(I_2^a+I_2^b)$ which yields to $I=\pi^2/2a+log(a)-1+2\big(-1+\int_{0}^{1} \frac{dx}{x} (\frac{1}{e^{x}-1}-\frac{1}{x}+\frac{1}{2}) +\int_{1}^{\infty}\frac{dx}{x} \frac{1}{e^{ x}-1}\big)+O(a)$. Now, the bracket is evaluated by noting that it has well behaved integrals and it equals to the zeroth order term of $\zeta(s)\Gamma(s)=\int_{0}^{\infty}  dx \frac{x^{s-1}}{e^x-1}$ as $s \rightarrow 0$. But this expansion equals with $-1/2s+1/2(\gamma_E-\ln(2 \pi))+O(s)$ and this completes the proof.
%
\renewcommand{\theequation}{B\arabic{equation}}
\setcounter{equation}{0}
\section{A Brief Introduction to ${\cal N}=4$ $SUSY$}
\label{B}
\subsection {The Lagranian: from ${\cal N}=1$ superfields to component fields.}
We fix our notation by assigning $\lambda^a$ to the gaugino, $f^{abc}$ are the structure constants of the gauge group $SU(N)$ and so $a,b,c=1,2...N^2-1$ while $i,j,k=1,2,3$ are the three flavors of the fermions $\psi_j^a$. The fermions and the gaugino are left Weyl spinors. There is one gauge field and three colored complex scalars $\phi_i^a$ which match the fermionic degrees of freedom. 

The ${\cal N}=4$ $SUSY$ Lagranian with a non abelian group may be written compactly as
\begin{align}\label{LN4SF}
{\cal L}_{{\cal N}_4}={\cal L}_{SYM}+{\cal L}_{WZ_{YM}}+{\cal W}_{{\cal N}_4}
\end{align}
where the first two terms may be found in any supersymmetry textbook \cite{Wess:1992cp,Terning:2006bq}. These two terms always exist in any ${\cal N}=1$ $SYM$, they respect ${\cal N}=4$ $SYM$ trivially  and may be written easily in terms of ${\cal N}=1$ superfields (see first two terms of $(7.6)$ and exercise $7.$ of Ch. $VII$ in \cite{Wess:1992cp} or (2.134)-(2.136) of \cite{Terning:2006bq}) 
while the last term is a particular superpotential (given below) which respects ${\cal N}=4$ in a non trivial way. In particular, ${\cal L}_{SYM}$ contains the first two terms of (\ref{LN4}) (see below) and in addition the bosonic auxiliary field $D^a$ in the form $1/2 D^a D^a$. The ${\cal L}_{WZ_{YM}}$ contains the gauge interactions of the $\psi$'s and $\phi$'s, the $\lambda \psi \phi$ interactions (see third, fourth and fifth terms of (\ref{LN4}) respectively), the term $igf^{abc} D^b \phi_i^{\dag a}  \phi_i^c$ and also the combination $F_i^aF_i^{\dag a}$ where $F_i^a$ is another auxiliary field. Finally, ${\cal W}_{{\cal N}_4}$ written in terms of ${\cal N}=1$ superfields $\Phi_i^a$ (see equation (14.32) of \cite{Terning:2006bq}) yields
\begin{align}\label{W4}
{\cal W}_{{\cal N}_4}=\frac{g}{3 \sqrt{2}} \epsilon_{ijk}f^{abc} \Phi_i^a\Phi_j^b\Phi_k^c\big|_{\theta \theta} + h.c.
=
\frac{g}{\sqrt{2}}\epsilon_{ijk}f^{abc}  \left(F_i^a \phi_j^b \phi_k^c - \psi_i^a \psi_j^b \phi_k^c \right) +h.c.
\end{align}
where in the second equality we have expanded out the $\Phi$'s according to equations (5.3) and (5.8) of \cite{Wess:1992cp}. Thus the ${\cal W}_{{\cal N}_4}$ contribution gives the sixth term of (\ref{LN4}) and one additional term containing the auxiliary field $F_i^a$. The remaining $\phi^4$ terms of (\ref{LN4}) are obtained by eliminating the two auxiliary fields in terms of the scalars setting the action on shell yielding to (14.33) of \cite{Terning:2006bq}\footnote{We choose the standard conventions for the fermions as in \cite{Dreiner:2008tw,Wess:1992cp} rather the convention of \cite{Terning:2006bq} used for (14.33).}
\begin{align}\label{LN4}
{\cal L}_{{\cal N}_4}=&-\frac{1}{4}F_{\mu \nu}F^{\mu \nu}-i\lambda^{\dag a}{\bar \sigma}^{\mu}D_{\mu}\lambda^a-i\psi_i^{a\dag} {\bar \sigma}^{\mu}D_{\mu}\psi_i^a+D^{\mu} \phi^{\dag}D^{\mu} \phi \notag\\&
-g \sqrt{2}  f^{abc}\left( \phi_i^{ \dag c} \lambda^a \psi_i^b-\psi_i^{\dag c} \lambda ^{\dag a} \phi_i^b \right)
-\frac{g}{\sqrt{2}}\epsilon_{ijk}f^{abc} \left( \phi_i^c \psi_j^a\psi_k^b+ \psi_i^{\dag c}\psi_j^{\dag a} \phi_k^{ \dag b} \right)\notag\\&
+\frac{g^2}{2}(f^{abc} \phi_i^b \phi_i^{\dag c}) (f^{ade} \phi_j^d \phi_j^{\dag e})
-\frac{g^2}{2}\epsilon_{ijk}\epsilon_{ilm}(f^{abc} \phi_j^b \phi_k^c)(f^{ade} \phi_l^{\dag d} \phi_m^{\dag e}).
\end{align}
%
%
%
\subsection{The vanishing of the beta function.}
It is known that the beta function for an $SU(N)$ gauge theory (for positive charge $g$) is given by
\begin{align}\label{beta}
\beta(g)=-\frac{1}{(4 \pi)^2} g^3 \left(\frac{11}{3}N-\frac{4}{3}\frac{1}{2}n_f \right)
\end{align}
where $n_f$ is the number of Dirac fermions while the factor of $1/2$ comes from the trace of two generators of the group. In the ${\cal N}=4$ $SUSY$ case, one should replace the $1/2$ by $N$ as the trace takes place in the adjoint representation and set $n_f=1/2 \times (3+1)=2$ for the $4$ (three matter fermions and one gaugino) Weyl spinors. On the other hand, the contribution of the scalars to the gauge boson self energy should be included as well.\footnote{We note that there are no additional corrections to the self energy of the fermions as the additional interactions that exist in the ${ \cal N}=4$ theory concerning the fermions, would change either completely the fermion ((see fourth term of (\ref{LN4}))) or the flavor and the color (see fifth term of (\ref{LN4})).}
 This contribution is due to the diagrams of figure \ref{pis} where for the $QED$ case yields to a combined contribution of $1/(48 \pi^2) g^3$ \cite{Srednicki:2007qs} and as a result, for $3 \times N$ (flavor$\times$colors) scalars, the two diagrams give the contribution  $1/(4 \pi)^2 3/3 g^3N$. Therefore, by adding all the contributions one gets
\begin{align}\label{beta1}
\beta(g)=-\frac{1}{(4 \pi)^2} g^3 \left(\frac{11}{3}N-\frac{4}{3} 2N -\frac{3}{3}N\right)=0,
\end{align}
that is the vanishing of the beta function. 
\subsection{Feynman rules at $T=0$}

The Feynman rules for two dimensional spinors may be found in \cite{Dreiner:2008tw} while the rest may be either derived from the Lagranian (\ref{LN4}) or they are known. As we work in TAG, there are no ghost propagators and vertices. The notation for the indices we use is
\begin{align}
i,j,k=1,2,3 \hspace{0.1in}\mbox{for flavor} \hspace{0.3in}  a,b,c=1,2,...,N^2-1 \hspace{0.1in}\mbox{for color} \hspace{0.3in} \alpha, {\dot \alpha}=1,2 \hspace{0.1in}\mbox{for spinors} 
\end{align}
%
%
\begin{figure}[!ht]
\centering
\includegraphics*[scale=0.75,angle=0,clip=true]{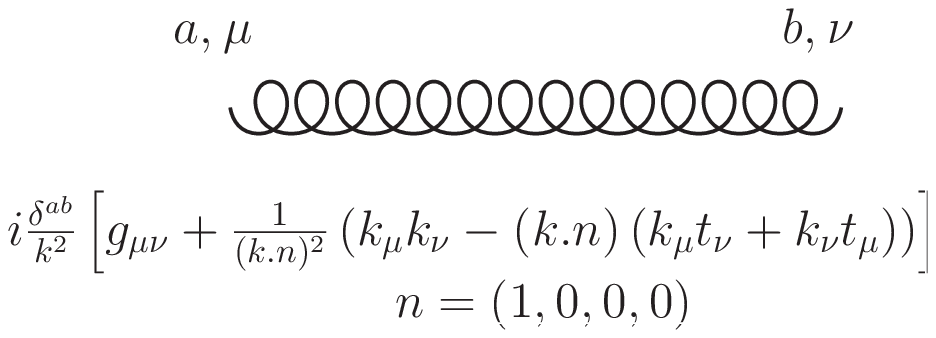} \hspace{1cm}
\includegraphics*[scale=0.75,angle=0,clip=true]{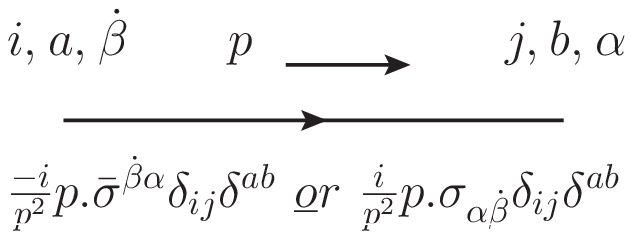}\vspace{1cm}
\includegraphics*[scale=0.75,angle=0,clip=true]{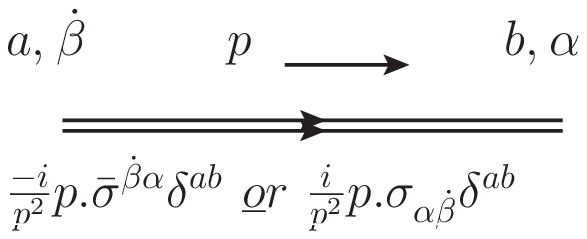} \hspace{3cm}
\includegraphics*[scale=0.75,angle=0,clip=true]{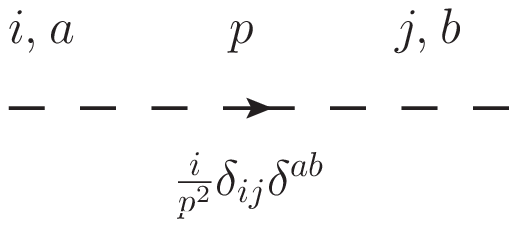}
\caption{Bare propagators for the gauge boson (up left) which is in TAG gauge, the fermions (up right), the gaugino (down left) and the scalars (down right). The choice of the appropriate rule for the fermions (and the gaugino) is determined by the the rest of the diagram where the propagators are attached to.}
\label{propFR}
\end{figure}
%
%
\begin{figure}[!th]
\centering
\includegraphics*[scale=0.65,angle=0,clip=true]{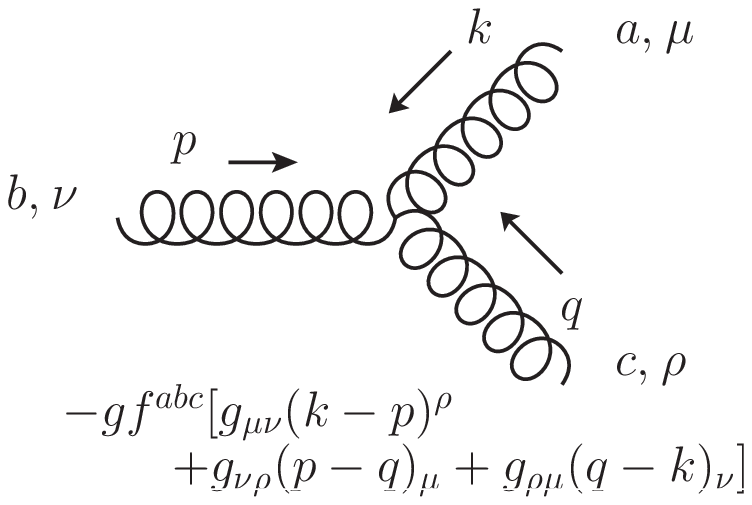} \hspace{0.5cm}
\includegraphics*[scale=0.65,angle=0,clip=true]{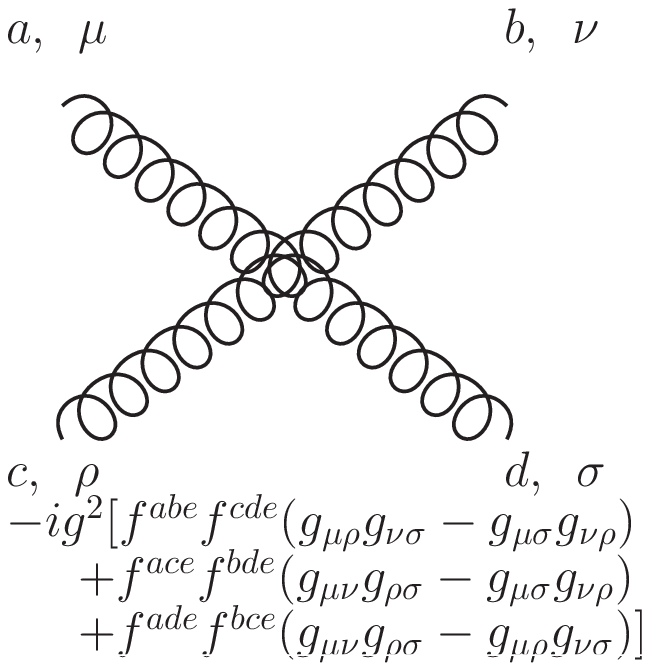}\hspace{0.5cm}
\includegraphics*[scale=0.65,angle=0,clip=true]{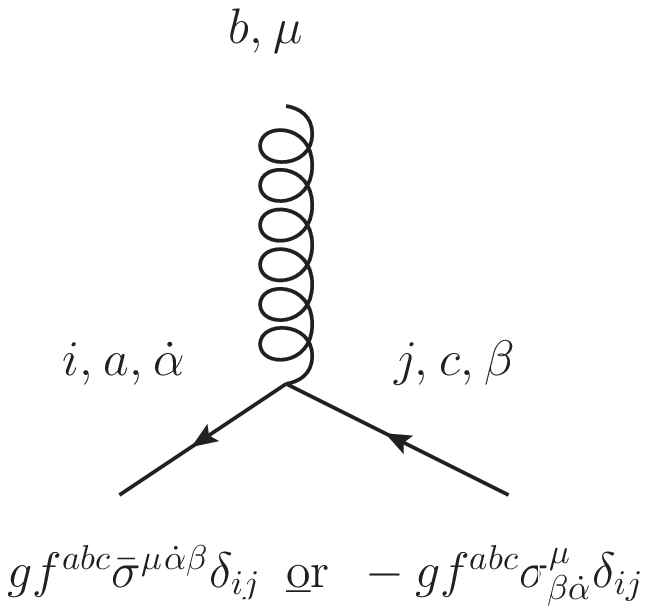} 
\caption{Gluon-gluon and gluon-fermion interactions (with negative charge $g$). For the gluon-fermion vertex a similar comment, as the one caption of figure 2 applies regarding the choice of the right rule. There also exists the analogue rule for the gaugino-gluon vertex without the $\delta_{ij}$.} 
\label{qcdvert}
\end{figure}
%
%
%
\begin{figure}[!th]
\centering
\hspace{0.5cm}\includegraphics*[scale=0.65,angle=0,clip=true]{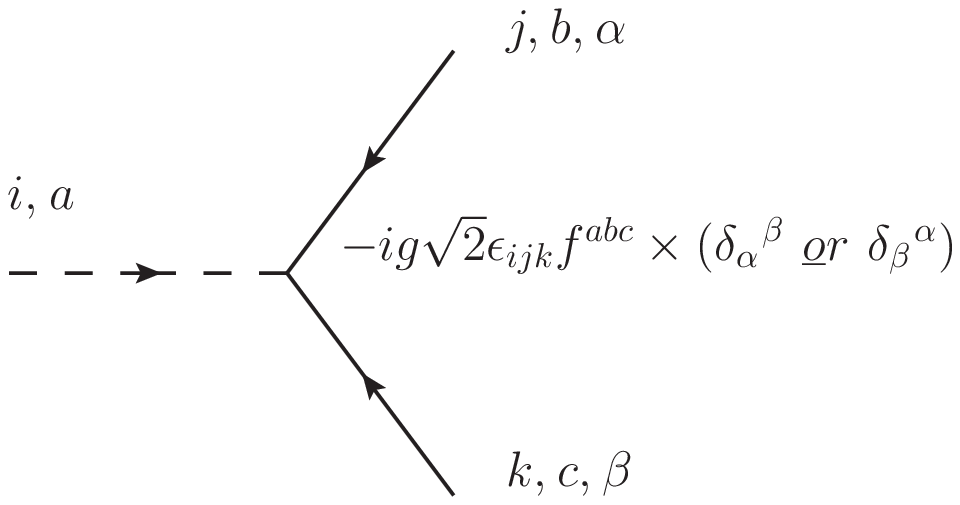} \hspace{1cm}
\includegraphics*[scale=0.65,angle=0,clip=true]{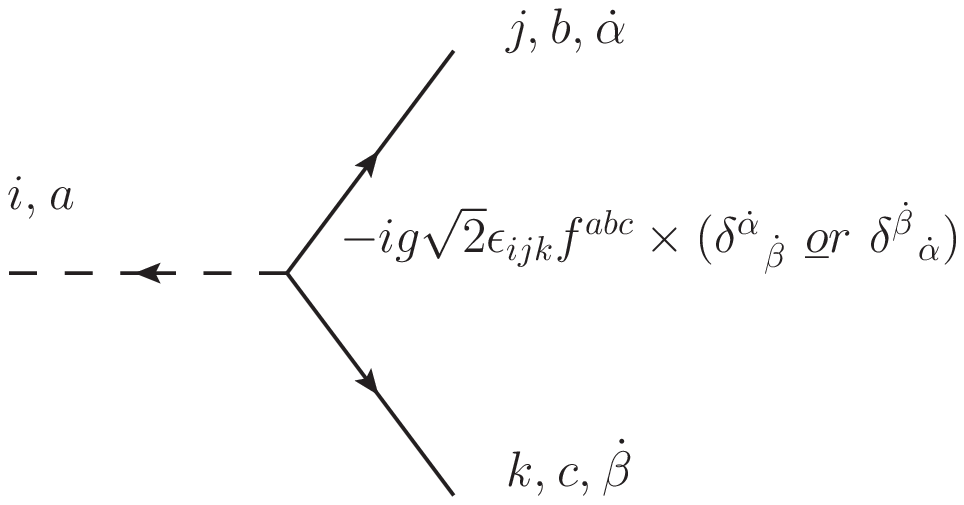} \vspace{1.5cm}
\includegraphics*[scale=0.65,angle=0,clip=true]{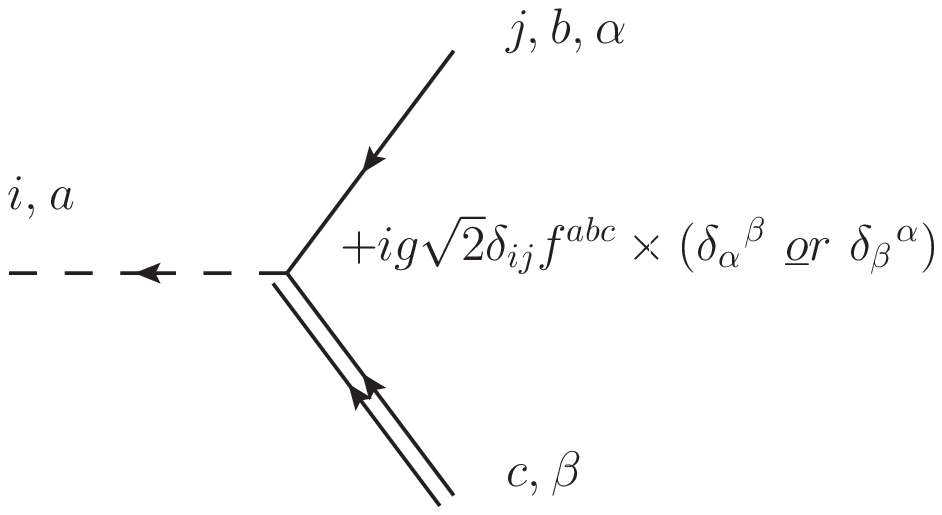} \hspace{1cm}
\includegraphics*[scale=0.65,angle=0,clip=true]{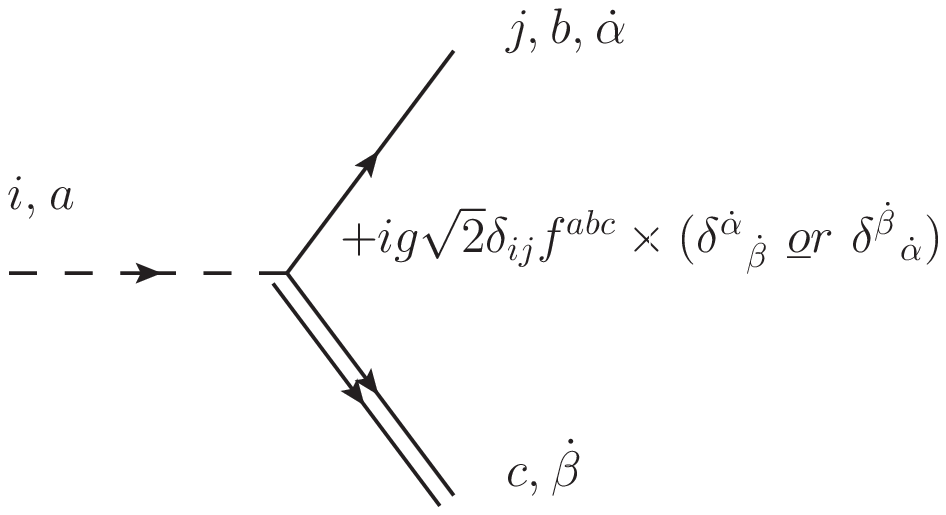}
\vspace{-1.5cm}
\caption{Yukawa vertices for fermion-fermion and fermion-gaugino with a similar comment as in the caption of figure 2 regarding the choice of the rule.}
\label{Yuk}
\end{figure}
%
%
\begin{figure}[!th]
\centering
\includegraphics*[scale=0.65,angle=0,clip=true]{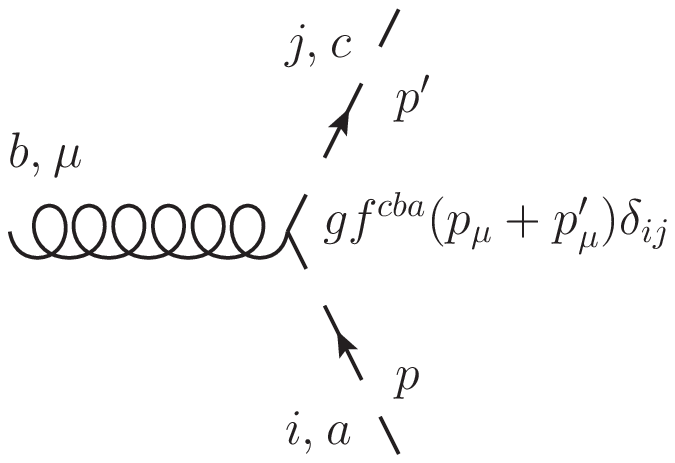} \hspace{0.5cm}
\includegraphics*[scale=0.65,angle=0,clip=true]{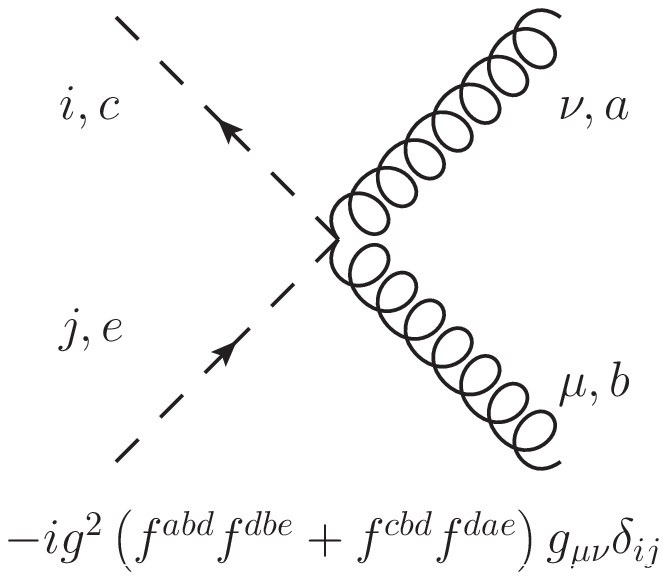}\vspace{0.5cm}
\includegraphics*[scale=0.65,angle=0,clip=true]{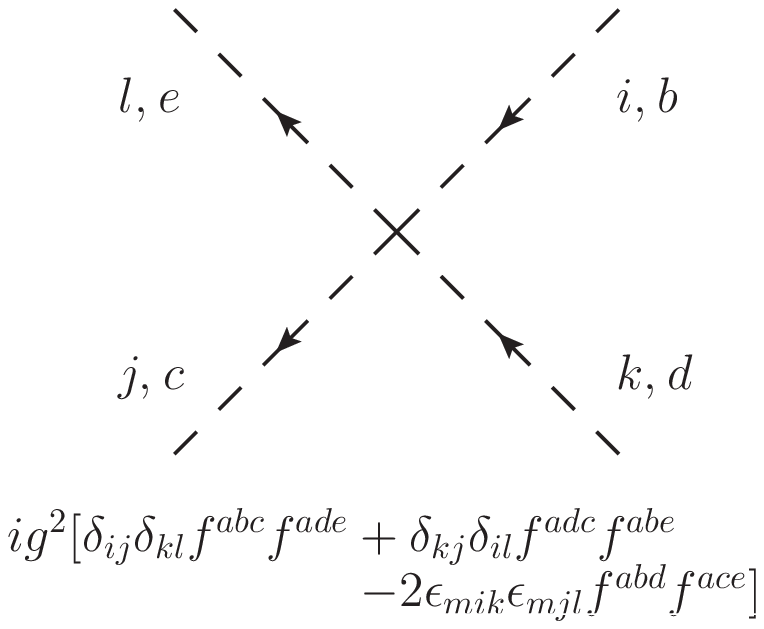} 
\caption{Scalar-gluon and scalar-scalar interactions.}
\label{gsFR}
\end{figure}
\subsection{Feynman rules at finite $T$}
They are obtained from those of $T=0$ using the following empirical rule: by multiplying all of the rules for $T=0$ by $(-i)$ and by including an additional minus sign to the gluon propagator (figure \ref{propFR} up-left), and the gluon-gluon interactions (left and middle diagrams of figure \ref{qcdvert}). In addition the following modifications are required
\begin{align}
\int \frac{d^4p}{(2 \pi)^4} \rightarrow  T \sum_n \int \frac{d^3{\bf p}}{(2 \pi)^3},  \hspace{0.4in} p_0 \rightarrow i2\pi T \times  \mbox{$n$ (bosons) or $(2n+1)$ (fermions), } n \in Z.
\end{align}
 %
 \vspace{-1cm}
\renewcommand{\theequation}{C\arabic{equation}}
\setcounter{equation}{0}
\section{Diagrams for the Dressed Propagator to $O(g^2)$}
\label{C}
There are six diagrams to $O(g^2)$ that contribute and they all involve the gluon propagator as, according to (\ref{LN4}), the scalar vertices change the flavor of the out-coming fermions and hence the related diagrams do not contribute.
%
\begin{figure}[!th]
\centering
\includegraphics*[scale=0.6,angle=0,clip=true]{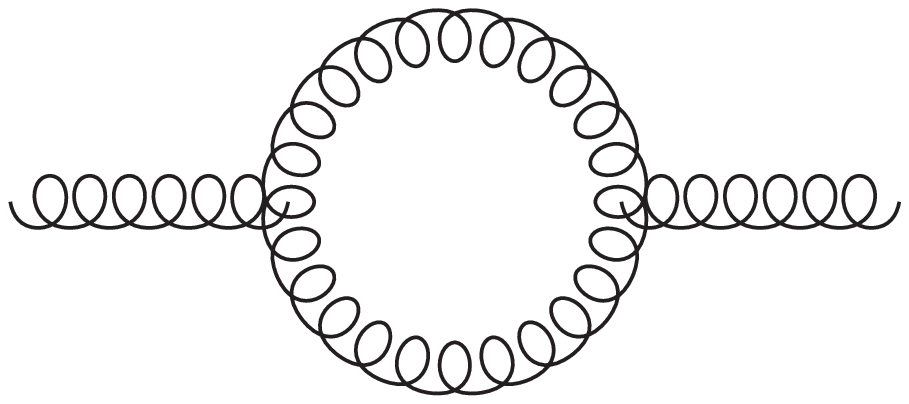} \hspace{1cm}
\includegraphics*[scale=0.6,angle=0,clip=true]{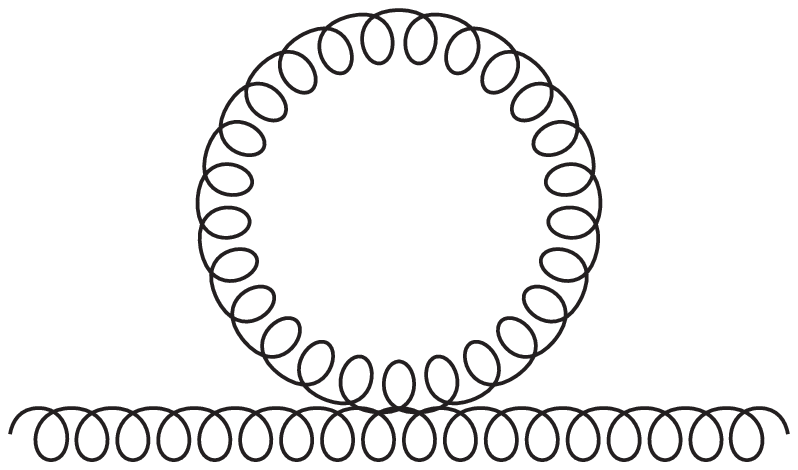}
\caption{Self corrections of the gluon propagator}
\label{pig}
\end{figure}
\begin{figure}[!th]
\centering
\includegraphics*[scale=0.55,angle=0,clip=true]{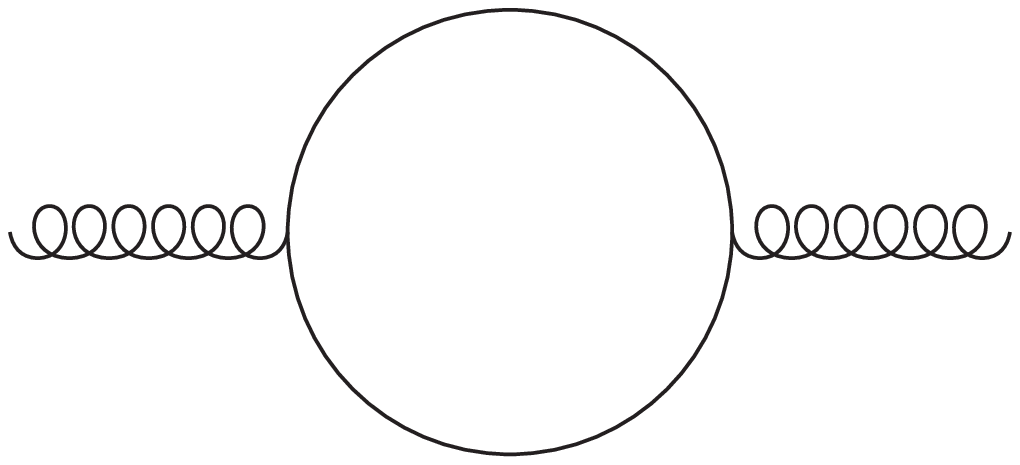} \hspace{1cm}
\includegraphics*[scale=0.55,angle=0,clip=true]{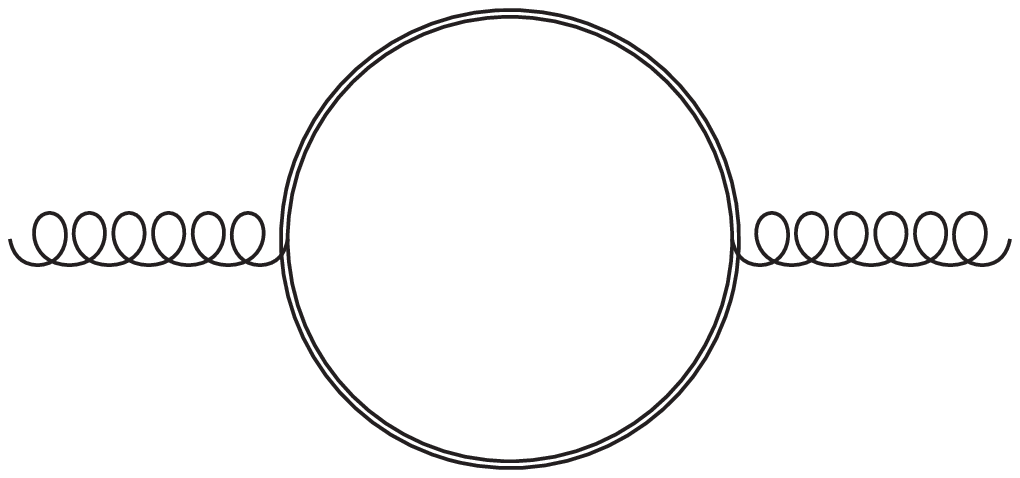}
\caption{Corrections due to the fermions (left) and the gauginos (right).}
\label{pif}
\end{figure}
\begin{figure}[!th]
\centering
\hspace{1cm} \includegraphics*[scale=0.6,angle=0,clip=true]{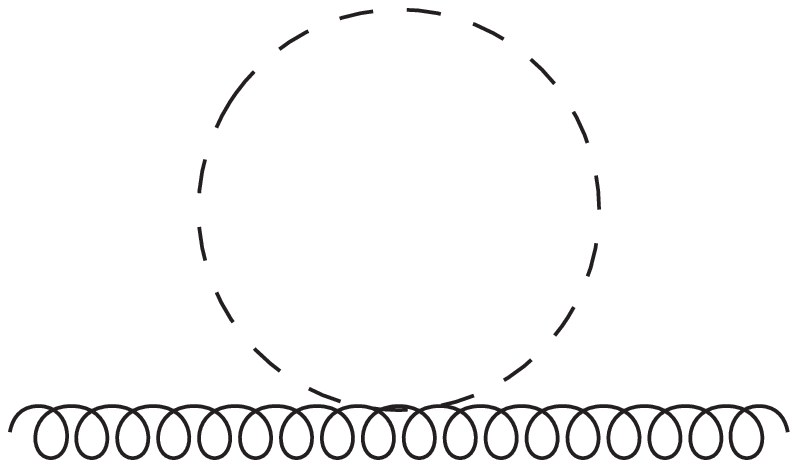} \hspace{1.5cm}
\includegraphics*[scale=0.6,angle=0,clip=true]{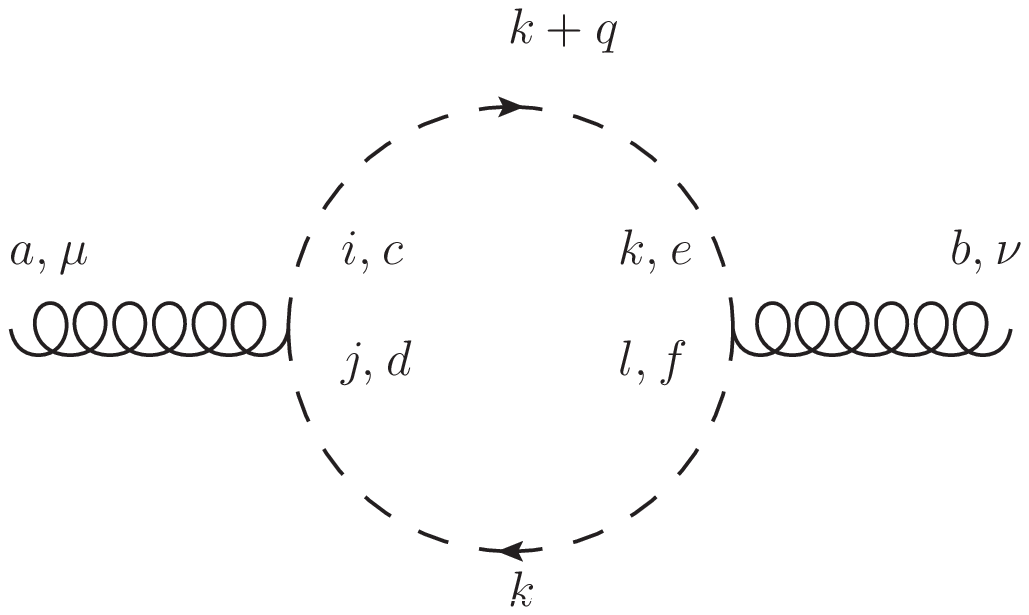}
\caption{Corrections due to the scalars.}
\label{pis}
\end{figure}
\newpage
\renewcommand{\theequation}{D\arabic{equation}}
\setcounter{equation}{0}
\section{Evaluating $\Pi_{s_2}^{00}$}
\label{D}

\begin{figure}[h!]
\centering
\hspace{1cm} \includegraphics*[scale=0.6,angle=0,clip=true]{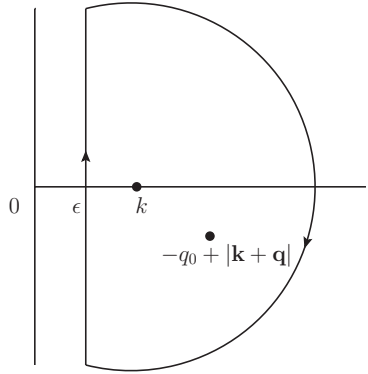} 
\caption{Contour of integration in the ${\tilde k}^0$ complex plane.}
\label{cont}
\end{figure}

Using the appropriate Feynman Rules of figures \ref{propFR} and \ref{gsFR} the right diagram of figure \ref{pis} reads
\begin{subequations}
\begin{align}
\Pi_{s_2}^{00 \hspace{0.01in}ab}=g^2  \delta_{ii}  f^{cad} &  f^{fbe}\delta^{ce}\delta^{df}T \sum_n \int \frac{d^3 {\bf k}}{(2 \pi)^3}    \frac{(2k^0 + q^0)(-2k^0 - q^0)}{ ((k^0)^2-{\bf k}^2)((k^0+q^0)^2 - ({\bf k}+{\bf q}) ^2)} \label{s2a} \\&
k^0=i2 \pi Tn    \hspace{0.7in}q^0=i2 \pi T m  \hspace{0.2in} \mbox{$n,m$ $\epsilon$ $\mathbb{Z}$} . \label{qoko}
\end{align} 
\end{subequations}
The thermal part of the summation formula (\ref{sumfo}) for bosons at zero chemical potential may take the form 
\begin{align}\label{sumB}
T\sum_nf(i2\pi T n)=\frac{1}{2 \pi i} \int_{-i \infty +\epsilon}^{i \infty +\epsilon} dk^0 \left(f(k^0)+f(-k^0)\right) N_B(k^0)
\end{align} 
But the corresponding $f(k^0)$ in the case of (\ref{s2a}) may be taken to be the integrand which has the property that the operation $k^0 \rightarrow -k^0$ is equivalent to $q^0 \rightarrow -q^0$. This observation will reduce our operations by a factor of two as we will only deal with the $+k^0$ piece and then just add the $-q^0$ contribution. 
In order to perform the integration we use the the contour of figure \ref{cont} where there exist two simple poles. Using (\ref{sff}) and performing the sum (in practice the integrations), (\ref{s2a}) becomes
\begin{align}\label{sb}
\Pi_{s_2}^{00 \hspace{0.01in}ab} \big|_{+q^0} =3g^2N \delta^{ab}  \int \frac{d^3 {\bf k}}{(2 \pi)^3}  \left( \frac{(2k+q^0)^2  N_B(k) }{2k ((k+q^0)^2 - ({\bf k}+{\bf q}) ^2)}   
 +\frac{   (2|{\bf k}+{\bf q}| -q^0)^2    N_B(-q^0+|{\bf k}+{\bf q}|)   }  {2\left((-q^0+|{\bf k}+{\bf q}|)^2 -k^2\right)  |{\bf k}+{\bf q}|  }   \right) 
\end{align}
where $k \equiv |{\bf k}|$. Now, using the property that $N_B(x+i 2\pi n)=N_B(x)$ $\forall$  $x$ and changing variables in the second piece  of the integrand from ${\bf k}$ to $-{\bf k}-{\bf q}$ results in the symmetrization of the two integrands under $q^0 \leftrightarrow -q^0$
\begin{align}\label{sc}
\Pi_{s_2}^{00 \hspace{0.01in}ab}\big|_{+q^0} =3g^2N \delta^{ab}  \int \frac{d^3 {\bf k}}{(2 \pi)^3} \frac{ N_B(k)}{2k} \left(  \frac{(2k+q^0)^2 }{ ((k+q^0)^2 - ({\bf k}+{\bf q}) ^2)}   +   \frac{(2k-q^0)^2 }{ ((k-q^0)^2 - (-{\bf k}-{\bf q}) ^2)}    \right).
\end{align}
Finally taking into account the $-q^0$ contribution (effectively multiplying (\ref{sc}) by a factor of two) and performing the angular integrations yields
\begin{align}\label{sd}
\Pi_{s_2}^{00 \hspace{0.01in}ab} =3\frac{g^2}{4 \pi^2}N \delta^{ab} \mbox{Re}   \int _0^{\infty}  dk \frac{1}{|{\bf q}|} (2k + q^0)^2   \ln\left( \frac{R_{+s}(q^0)}{R_{-s}(q^0)}\right) N_B(k)
\end{align}
where the operator Re is defined in (\ref{Re}) while $R_{\pm s}=(q^0)^2-{\bf q}^2+2 k q^0\pm 2 k |{\bf q}|$ and $k \equiv |{\bf k}|$.



\bibliography{references}                   

\end{document}